\documentclass[conference,a4paper]{IEEEtran}
\IEEEoverridecommandlockouts
\usepackage{cite}
\usepackage{amsmath,amssymb,amsfonts}
\usepackage{algorithmic}
\usepackage{graphicx}
\usepackage{textcomp}
\usepackage{xcolor}
\usepackage{booktabs}
\usepackage{multirow}
\usepackage{array}
\usepackage[colorlinks=true,linkcolor=blue,citecolor=blue,urlcolor=magenta]{hyperref}
\usepackage{flushend}

\def\BibTeX{{\rm B\kern-.05em{\sc i\kern-.025em b}\kern-.08em
    T\kern-.1667em\lower.7ex\hbox{E}\kern-.125emX}}

\makeatletter
\def\blfootnote#1{%
  \begingroup
  \renewcommand\thefootnote{}\footnote{#1}%
  \addtocounter{footnote}{-1}%
  \endgroup
}
\makeatother

\begin{document}

\IEEEpubid{\makebox[\columnwidth]{979-8-3195-1464-6/26/\$31.00~\copyright~2026 IEEE \hfill}
\hspace{\columnsep}\makebox[\columnwidth]{}}

\title{Beyond Local Power: Functional Connectivity Analysis for Subject-Independent Learning Style Recognition}

\author{\relax}

\newcommand{\authorcol}[2]{%
  \begin{minipage}[t]{0.32\textwidth}
    \centering
    #1\\
    #2
  \end{minipage}%
}

\makeatletter
\def\@maketitle{%
  \newpage
  \null
  \vskip 2em
  \begin{center}%
  \let \footnote \thanks
    {\LARGE \@title \par}%
    \vskip 1.5em%
    {\large
      \authorcol{Wiga Maulana Baihaqi}{%
        \textit{Dept. of Electrical and Information Engineering, Faculty of Engineering}\\
        \textit{Universitas Gadjah Mada}\\
        Yogyakarta, Indonesia\\
        wiga.mti15@mail.ugm.ac.id}%
      \authorcol{Indriana Hidayah}{%
        \textit{Dept. of Electrical and Information Engineering, Faculty of Engineering}\\
        \textit{Universitas Gadjah Mada}\\
        Yogyakarta, Indonesia\\
        indriana.h@ugm.ac.id}%
      \authorcol{Sri Kusrohmaniah}{%
        \textit{Department of Psychology, Faculty of Psychology}\\
        \textit{Universitas Gadjah Mada}\\
        Yogyakarta, Indonesia\\
        koes\_psi@ugm.ac.id}%

      \vskip 1.2em

      \authorcol{Noor Akhmad Setiawan}{%
        \textit{Dept. of Electrical and Information Engineering, Faculty of Engineering}\\
        \textit{Universitas Gadjah Mada}\\
        Yogyakarta, Indonesia\\
        noorwewe@ugm.ac.id}%
    }
  \end{center}%
  \par
  \vskip 1.5em}
\makeatother

\maketitle

\blfootnote{\copyright~2026 IEEE. Personal use of this material is permitted. Permission from IEEE must be obtained for all other uses, in any current or future media, including reprinting/republishing this material for advertising or promotional purposes, creating new collective works, for resale or redistribution to servers or lists, or reuse of any copyrighted component of this work in other works.}

\begin{abstract}
Identifying individual learning styles optimizes pedagogical efficacy. While traditional questionnaires are structured, behavioral tracking methods require prolonged interaction log accumulation. To overcome these temporal constraints, this paper proposes an objective Electroencephalography (EEG) approach evaluating Phase Locking Value (PLV) connectivity against localized features across the Active--Reflective (AR) and Verbal--Visual (VV) Felder-Silverman dimensions. EEG signals were recorded from 28 participants during Raven's Advanced Progressive Matrices tasks. Support Vector Machine classification used Leave-One-Subject-Out Cross-Validation (LOSO-CV) alongside a 70:30 intra-subject split. The VV dimension achieved 70.00\% subject-level accuracy driven by distinct fronto-occipital polarization. Conversely, the AR dimension yielded lower cross-subject generalizability (55.56\%) due to overlapping executive networks and a ``Systematic Neural Inversion'' phenomenon, where stable individual connectivity signatures operated diametrically opposed to global boundaries (up to 20--0 voting margins). Ultimately, these outcomes demonstrate that rigid ``one-size-fits-all'' classifiers are bounded by biological diversity, emphasizing the need for future adaptive feature transformation techniques to bridge the cross-subject generalization gap.
\end{abstract}


\IEEEpubidadjcol

\section{Introduction}

Identifying individual learning styles is essential for optimizing overall pedagogical efficacy and enhancing the broader student learning experience \cite{ref1}. While traditional self-report questionnaires like the Felder-Silverman Index of Learning Styles (ILS) offer a structured \cite{ref2}, widely adopted approach for identifying learning tendencies \cite{ref3}, alternative data-driven methods have emerged to infer learning styles from behavioral patterns or log data, aiming to eliminate the need for lengthy self-reports. However, these behavioral classifications are inherently time-consuming, often requiring multiple sessions across an entire course to accumulate sufficient interaction logs \cite{ref4}. To overcome these temporal constraints, Electroencephalography (EEG) provides a rapid, direct alternative by recording cortical activity with excellent temporal resolution \cite{ref5}, allowing the immediate capture of short-term neurophysiological dynamics during specific tasks. Recent advancements have shifted from simple spectral clustering to advanced machine and deep learning architectures \cite{ref6} utilizing time-domain, frequency-domain, and non-linear connectivity features \cite{ref7,ref8}.

Despite these advances, existing deep learning models heavily rely on localized power features that are sensitive to anatomical differences between subjects, leading to poor generalization across different populations \cite{ref9}. Furthermore, localized approaches treat brain regions in isolation, failing to account for the synchronized, large-scale networks that underpin complex learning activities. Neuroscientific evidence suggests that cognitive functions---such as information processing (Active--Reflective) or mental imagery (Verbal--Visual)---emerge from the cooperation of distributed neural networks \cite{ref10}. Recent efforts have also explored adaptive wavelet-based decomposition and multi-instance learning strategies to further improve EEG-based learning style recognition \cite{ref11}. Specifically, phase synchronization between fronto-parietal and fronto-occipital circuits facilitates cognitive state switching. Phase Locking Value (PLV) effectively measures this synchrony independent of signal amplitude, mitigating individual physiological variations and providing neurophysiologically well-founded spatial priors that enhance classification accuracy and model interpretability \cite{ref9}.

This paper evaluates PLV against traditional localized features (PSD, entropy, and statistical measures) for FSLSM classification. Using Support Vector Machines (SVM), we classify the Active--Reflective and Verbal--Visual dimensions. To ensure robust generalization, we employ a dual-validation strategy: a standard 70:30 train-test split and a rigorous Leave-One-Subject-Out Cross-Validation (LOSO-CV) to assess subject-independent performance.

The contributions of this paper are threefold. First, it provides a complete feature evaluation across the frequency, time, complexity, and connection domains to discover highly accurate EEG biomarkers for the Felder-Silverman Learning Style Model (FSLSM). Second, to address the underlying neuronal heterogeneity in EEG data, this study establishes a subject-independent generalization framework applying a rigorous Leave-One-Subject-Out Cross-Validation (LOSO-CV) methodology. Finally, we conduct a full neurophysiological characterisation by directly mapping the FSLSM dimensions to phase-locking brain networks to understand the neurophysiological foundation underpinning individual learning styles.

The remainder of this paper is organized as follows: Section II details the methodology; Section III reports the results; Section IV discusses the findings; and Section V concludes the paper.

\section{Method}

\begin{figure*}[!t]
\centering
\includegraphics[width=\textwidth]{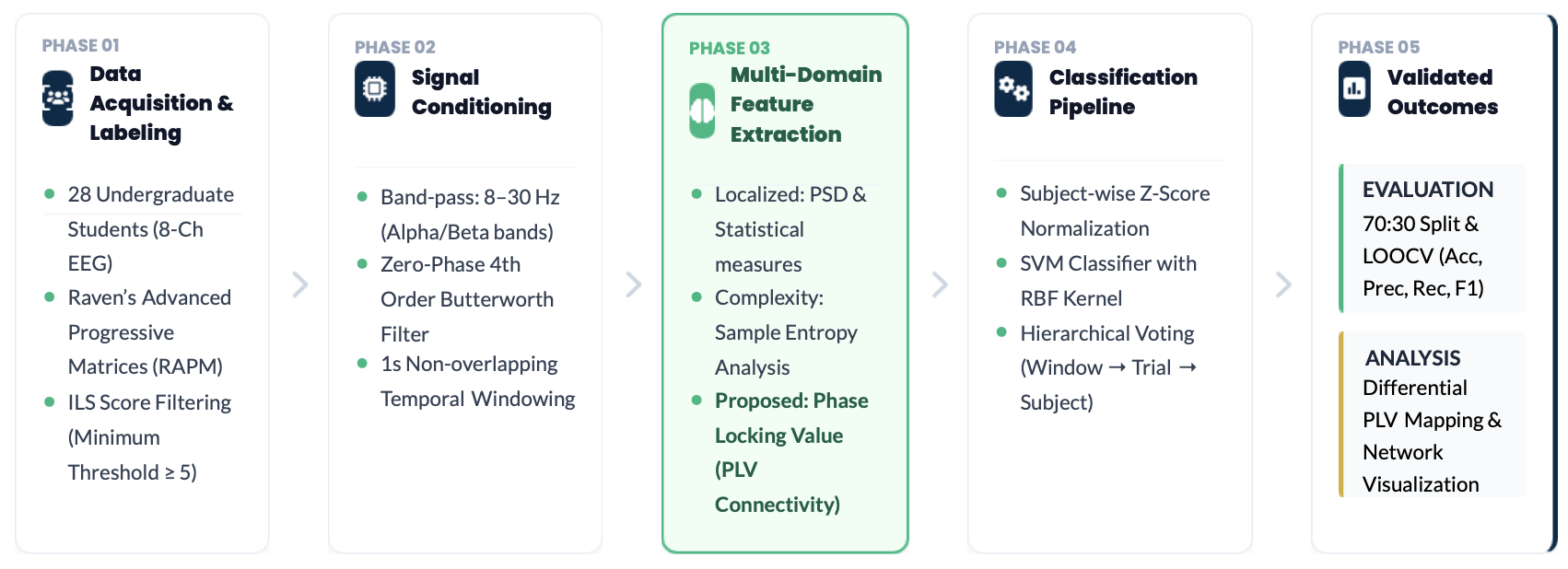}
\caption{Proposed method: step-wise EEG-based FSLSM recognition pipeline.}
\label{fig1}
\end{figure*}

The step-wise scheme of the suggested research methodology for identifying the FSLSM facets using EEG data is illustrated in Fig.~\ref{fig1}. The research has been divided into five parts: (1) recording and labeling EEG signals while participants were engaged in mental tasks, (2) preprocessing signals, (3) extracting features from different domains, (4) classifying signals using machine learning, and (5) evaluating the results and performing neurophysiological analysis. Through this extensive methodology, we intend to transmute the raw brainwave data into steady cognitive profiles while at the same time addressing the brain signal individual variability issue. Every step in the process, together with the specific computational and validation methods, is described in the following subsections.

\subsection{Participants}

Twenty-eight healthy college students (aged 18--20) with no history of neurological or psychiatric disorders participated in this study. To ensure robust labeling and minimize ``neural inversion'' artifacts \cite{ref6}, only participants with a score of $\geq 5$ (moderate-to-strong preference) on the Index of Learning Styles (ILS) survey were included; balanced learners (scores 1--3) were excluded. Following a balanced-class approach \cite{ref6}, we selected 18 participants for the Active--Reflective (AR) dimension (9 Active, 9 Reflective) and 10 for the Verbal--Visual (VV) dimension (5 Verbal, 5 Visual). This 50:50 distribution establishes a strict theoretical chance-level accuracy of 50.00\%, eliminating majority class bias. The study was approved by the Research Ethics Committee of Universitas Indonesia Maju (No: 2649/Sket/Ka-Dept/RE/UIMA/VI/2025), and all participants provided written informed consent.

\subsection{EEG Data Acquisition}

\begin{figure}[!t]
\centering
\includegraphics[width=0.85\columnwidth]{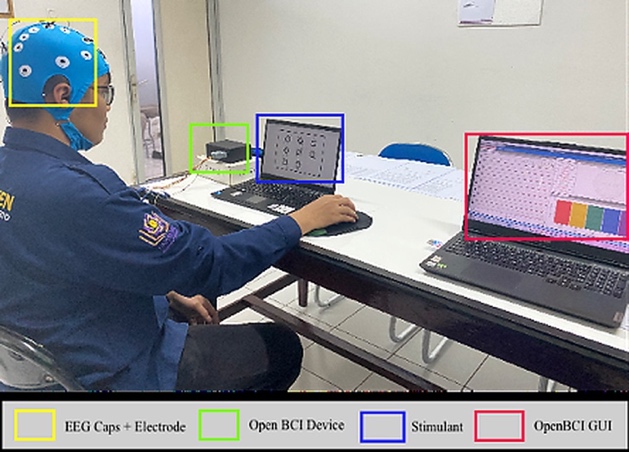}
\caption{Experimental setup for EEG data acquisition.}
\label{fig2}
\end{figure}

Continuous EEG signals were recorded at a 250~Hz sampling rate using an OpenBCI Cyton board from eight electrodes (Fp1, Fp2, F3, F4, C3, C4, O1, and O2) with a left earlobe (A1) reference \cite{ref12}. Cognitive stimulation was induced via 20 independent, 15-second trials of Raven's Advanced Progressive Matrices (RAPM) Set II (Fig.~\ref{fig2}). RAPM serves as a language-free, high-load cognitive catalyst that forces style-specific processing strategies under the FSLSM framework. Specifically, the logical symbol rules within RAPM effectively elicit divergent processing behaviors: reflective learners naturally engage in deliberative introspection and meticulous hypothesis evaluation, whereas active learners instinctively deploy rapid, intuitive mental trial-and-error operations \cite{ref6}. Simultaneously, solving these non-verbal matrices forces distinct input decoding strategies, where visual learners instinctively rely on spatial imagery and pattern manipulation, while verbal learners formulate internal sequential logical rules to decode the matrix structures \cite{ref13}. Furthermore, the moderate difficulty level of RAPM prevents cognitive fatigue and excessive workload, thereby ensuring high signal integrity while capturing these distinct functional connectivity topologies. Fig.~\ref{fig2} illustrates the experimental setup, including the participant cap, Cyton board, and real-time monitoring via the OpenBCI GUI. The total recording time was approximately five minutes per participant.

\subsection{Preprocessing}

To enhance signal quality while preserving phase information, raw EEG data were band-pass filtered between 8--30~Hz using a fourth-order, zero-phase Butterworth filter \cite{ref14}. The 15-second trials were segmented into 1-second non-overlapping windows. To ensure reliable functional connectivity estimation, features were calculated at the whole-trial level by combining these windowed segments. In line with previous protocols, no additional artifact rejection or re-referencing was performed to maintain methodological consistency.

\subsection{Feature Extraction}

Four feature sets were extracted across the time, frequency, complexity, and connectivity domains:
\begin{itemize}
\item Power Spectral Density (PSD) via Welch's method for Alpha (8--13~Hz) and Beta (13--30~Hz) bands across 8 channels (16 features).
\item Five statistical measures (mean, standard deviation, skewness, kurtosis, and RMS) per channel (40 features).
\item Sample Entropy (SE) per channel to index cognitive load (8 features).
\item Phase Locking Value (PLV) to quantify phase synchronization consistency between channel pairs.
\end{itemize}

The PLV of two channels $i$ and $j$ is given by (\ref{eq:plv}) \cite{ref15}:
\begin{equation}
PLV_{i,j} = \left| \frac{1}{N}\sum_{n=1}^{N} e^{j\left(\phi_i(n) - \phi_j(n)\right)} \right|
\label{eq:plv}
\end{equation}
where $\phi(n)$ is the instantaneous phase acquired by the Hilbert Transform and $N$ stands for the number of samples. For the 8-channel system, 28 unique pairs were computed per window, representing the global synchronization of the brain network.

\subsection{Classification and Evaluation Framework}

To address inter-subject heterogeneity, individual Z-score normalization was applied to each participant's data. Classification was performed using a Support Vector Machine (SVM) with a Radial Basis Function (RBF) kernel \cite{ref16}. The SVM decision function for an input $\mathbf{x}$ is defined as (\ref{eq:svm}):
\begin{equation}
f(\mathbf{x}) = \text{sign}\left( \sum_{k=1}^{n_{sv}} \alpha_k y_k K(\mathbf{x}_k, \mathbf{x}) + b \right)
\label{eq:svm}
\end{equation}
here, $n_{sv}$ is the number of support vectors, $\alpha_k$ are the Lagrange multipliers, $y_k$ are the class labels, $b$ is the bias term, and $K(\mathbf{x}_k, \mathbf{x})$ is the RBF kernel.

Subject-level predictions were determined using a hierarchical two-stage majority voting system across windows and trials. Model generalization was evaluated using two distinct validation schemes: an epoch-wise 70:30 split and Leave-One-Subject-Out Cross-Validation (LOSO-CV). The epoch-wise 70:30 split was conducted as an internal validation baseline to confirm model learnability and intra-subject feature consistency. Meanwhile, LOSO-CV was implemented to rigorously evaluate subject-independent performance while completely eliminating data leakage risks; notably, all core findings and subject-independent claims in this study rely exclusively on the LOSO-CV scheme.

Performance across both schemes was quantified using Accuracy, Precision, Recall, and F1-Score for each FSLSM dimension.

\subsection{Neurophysiological Connectivity Analysis}

Functional connectivity was estimated by calculating the Phase Locking Value (PLV) between pairs of electrodes among a set of eight electrodes located at frontal (Fp1, Fp2, F3, F4), central (C3, C4), and occipital (O1, O2) sites. Class-specific connectivity patterns were obtained by retaining only the top 25\% strongest synchronizations for each class. To detect features that best discriminate between classes, a difference connectivity matrix ($\Delta$PLV) was derived through a double threshold technique \cite{ref17}. Edges were displayed only when they satisfied two requirements: (1) statistical significance by means of an independent t-test ($p < 0.05$), and (2) a size among the top 20\% of absolute differences ($\Delta$PLV $>$ 80th percentile). Differential edges were assigned color codes (red for positive, blue for negative differences) and thicknesses proportional to their $\Delta$PLV values. With this, the resulting networks represent statistically supported and neurophysiologically significant features of FSLSM.

\section{Experimental Results}

\subsection{Validation of the Feature Extraction Pipeline}

\begin{figure*}[!t]
\centering
\includegraphics[width=\textwidth]{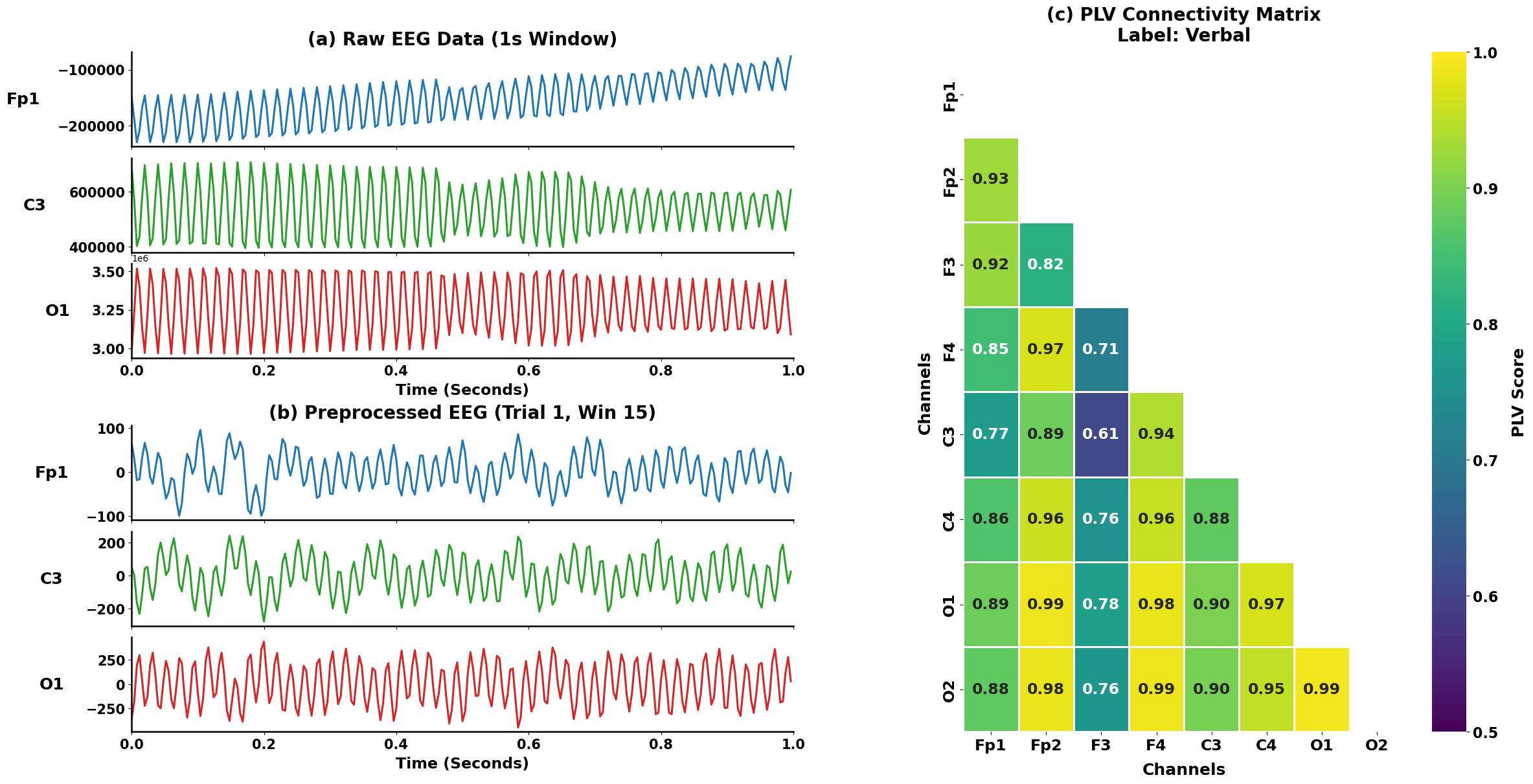}
\caption{EEG data processing pipeline showing (a) raw signals, (b) 8--30 Hz filtered signals, and (c) the derived PLV matrix.}
\label{fig3}
\end{figure*}

The signal transformation pipeline is illustrated in Fig.~\ref{fig3}. Panel (a) shows the raw EEG data from representative pre-frontal (Fp1), central (C3), and occipital (O1) channels, which exhibit substantial DC offset and low-frequency baseline drifts. Panel (b) confirms the efficacy of the 8--30~Hz Butterworth band-pass filter, which transforms the raw data into stable, zero-mean signals within normal neurophysiological ranges by isolating Alpha and Beta oscillations. Finally, Panel (c) displays the derived Phase Locking Value (PLV) connectivity matrix for a Verbal learner. This heatmap captures distinct, noise-free spatial synchronization topologies that serve as the structural features for FSLSM classification.

\subsection{Comparative Performance of EEG Feature Domains}

We evaluated classification performance across four domains: Phase Locking Value (PLV), Power Spectral Density (PSD), Sample Entropy, and Time-domain Statistics. Table~\ref{tab1} presents the Leave-One-Subject-Out (LOSO) cross-validation results, testing the model's ability to generalize to unseen subjects.

\begin{table}[!t]
\caption{Subject-Level Classification Performance: LOSO Validation}
\label{tab1}
\centering
\footnotesize
\resizebox{\columnwidth}{!}{%
\begin{tabular}{@{}llcccc@{}}
\toprule
\textbf{Dim.} & \textbf{Feature Domain} & \textbf{Acc} & \textbf{Prec} & \textbf{Rec} & \textbf{F1} \\
\midrule
\multirow{4}{*}{AR (Act--Ref)}
 & PLV               & 55.56\% & 55.84\% & 55.56\% & 55.00\% \\
 & Entropy           & 55.56\% & 56.92\% & 55.56\% & 53.25\% \\
 & PSD (Alpha/Beta)  & 61.11\% & 61.25\% & 61.11\% & 60.99\% \\
 & Statistics        & 55.56\% & 56.92\% & 55.56\% & 53.25\% \\
\midrule
\multirow{4}{*}{VV (Ver--Vis)}
 & PLV               & 70.00\% & 70.83\% & 70.00\% & 69.70\% \\
 & PSD               & 40.00\% & 40.00\% & 40.00\% & 40.00\% \\
 & Statistics        & 40.00\% & 40.00\% & 40.00\% & 40.00\% \\
 & Entropy           & 40.00\% & 40.00\% & 40.00\% & 40.00\% \\
\bottomrule
\end{tabular}%
}
\end{table}

As shown in Table~\ref{tab1}, PSD (Alpha/Beta) features yielded the highest accuracy for the Active--Reflective (AR) dimension (61.11\%), indicating that AR styles are better captured by rhythmic oscillations and spectral power distributions. Conversely, PLV achieved the highest accuracy for the Verbal--Visual (VV) dimension (70.00\%). This demonstrates that VV processing is predominantly a large-scale cortical network phenomenon relying on inter-regional synchronization (e.g., occipital-temporal integration) rather than localized power alterations.

\subsection{Analysis of the Generalization Gap (LOSO vs. 70:30 Split)}

A substantial ``generalization gap'' was observed when comparing the subject-independent LOSO scheme against the intra-subject 70:30 train-test split baseline (Table~\ref{tab2}).

\begin{table}[!t]
\caption{Subject-Level Performance: 70:30 Train-Test Split}
\label{tab2}
\centering
\footnotesize
\resizebox{\columnwidth}{!}{%
\begin{tabular}{@{}llcccc@{}}
\toprule
\textbf{Dimension} & \textbf{Feature Domain} & \textbf{Acc} & \textbf{Prec} & \textbf{Rec} & \textbf{F1} \\
\midrule
\multirow{2}{*}{AR (Active--Refl.)}
 & PLV & 100\%   & 100\%   & 100\%   & 100\% \\
 & PSD & 94.44\% & 95.00\% & 94.44\% & 94.43\% \\
\midrule
\multirow{2}{*}{VV (Verbal--Vis.)}
 & PLV        & 100\%   & 100\%   & 100\%   & 100\% \\
 & Statistics & 90.00\% & 91.67\% & 90.00\% & 89.90\% \\
\bottomrule
\end{tabular}%
}
\end{table}

In the 70:30 split, PLV achieved perfect accuracy (100.00\%) for both dimensions. However, the sharp performance decline in LOSO validation highlights pronounced inter-subject variability. While the SVM model effectively learns individual neural ``fingerprints'' from known subjects, cross-subject generalization remains highly challenging due to the idiosyncratic nature of EEG signals.

\subsection{Stability and Majority Voting Efficiency}

The hierarchical majority voting mechanism at the trial and subject levels effectively filtered out transient, noisy EEG segments to stabilize classification.

\begin{table}[!t]
\caption{Stability Metrics: Mean Acc $\pm$ SD (LOSO)}
\label{tab3}
\centering
\footnotesize
\begin{tabular}{@{}llccc@{}}
\toprule
\textbf{Dim.} & \textbf{Feat.} & \textbf{Window Acc} & \textbf{Trial Acc} & \textbf{Subj. Acc} \\
\midrule
AR & PSD & 55.89\% $\pm$ 25.11\% & 56.11\% $\pm$ 34.62\% & 61.11\% \\
VV & PLV & 55.47\% $\pm$ 31.03\% & 56.00\% $\pm$ 34.91\% & 70.00\% \\
\bottomrule
\end{tabular}
\end{table}

Table~\ref{tab3} demonstrates that aggregating window-level predictions into trial- and subject-level decisions consistently enhances classification stability. For instance, the AR subject-level accuracy (61.11\%) outperforms its window-level baseline (55.89\%). This indicates that while brief EEG epochs may be ambiguous, the cumulative majority vote over the entire session provides a more resilient reflection of a learner's cognitive style.

\subsection{Neurophysiological Connectivity Analysis}

Functional synchronization networks were visualized using PLV mapping to investigate the biological underpinnings of classification performance.

\subsubsection{Active--Reflective (AR) Connectivity Profiles}

\begin{figure*}[!t]
\centering
\includegraphics[width=\textwidth]{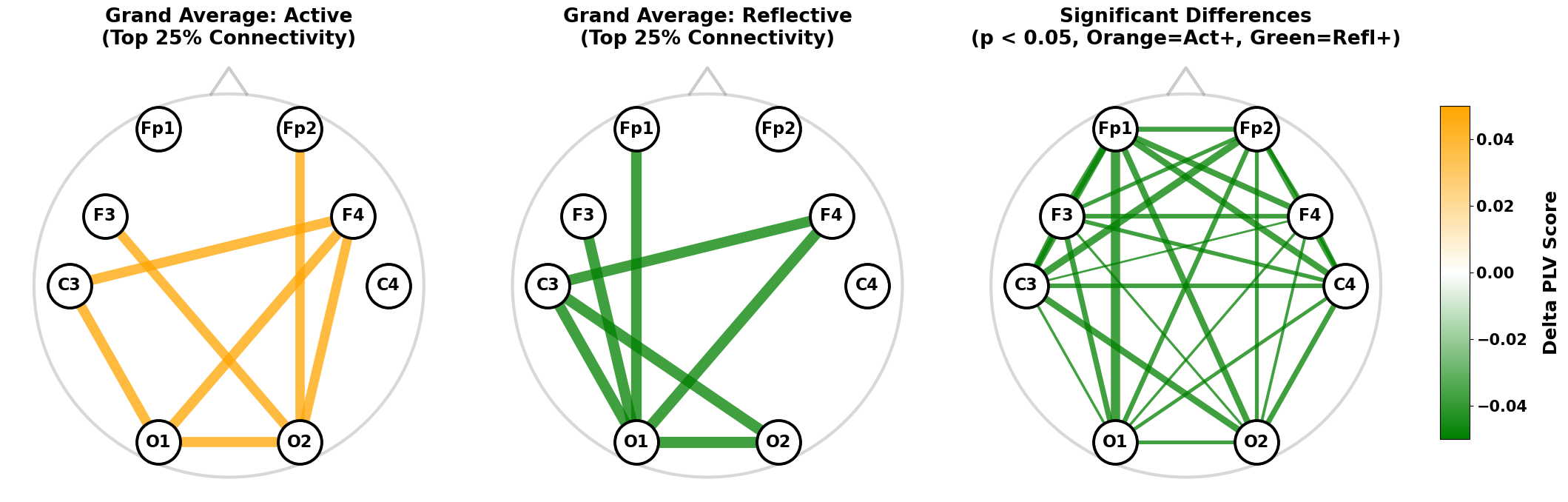}
\caption{Connectivity analysis for the AR dimension: (a) grand average Active, (b) grand average Reflective, and (c) significant differences.}
\label{fig4}
\end{figure*}

Connectivity patterns (Fig.~\ref{fig4}) reveal distinct synchronization densities. The Active style exhibits a streamlined network localized along the Frontal--Occipital axis (Fp2-O2, F4-O2). Conversely, the Reflective style displays a denser, more complex global integration across Frontal--Central--Occipital regions. The differential map (Fig.~\ref{fig4}c) is dominated by blue edges (Reflective $>$ Active), showing that the Reflective style recruits significantly higher global synchronization driven by intensive mental simulation and executive control \cite{ref17}. Key discriminative features include dense inter-hemispheric coupling (Fp1-Fp2, C3-C4) and robust long-range prefrontal-to-occipital connectivity. This topological overlap and shared executive circuitry explain why the classifier struggles to distinguish baseline synchronization profiles across unseen subjects during LOSO validation.

\subsubsection{Verbal--Visual (VV) Connectivity Profiles}

\begin{figure*}[!t]
\centering
\includegraphics[width=\textwidth]{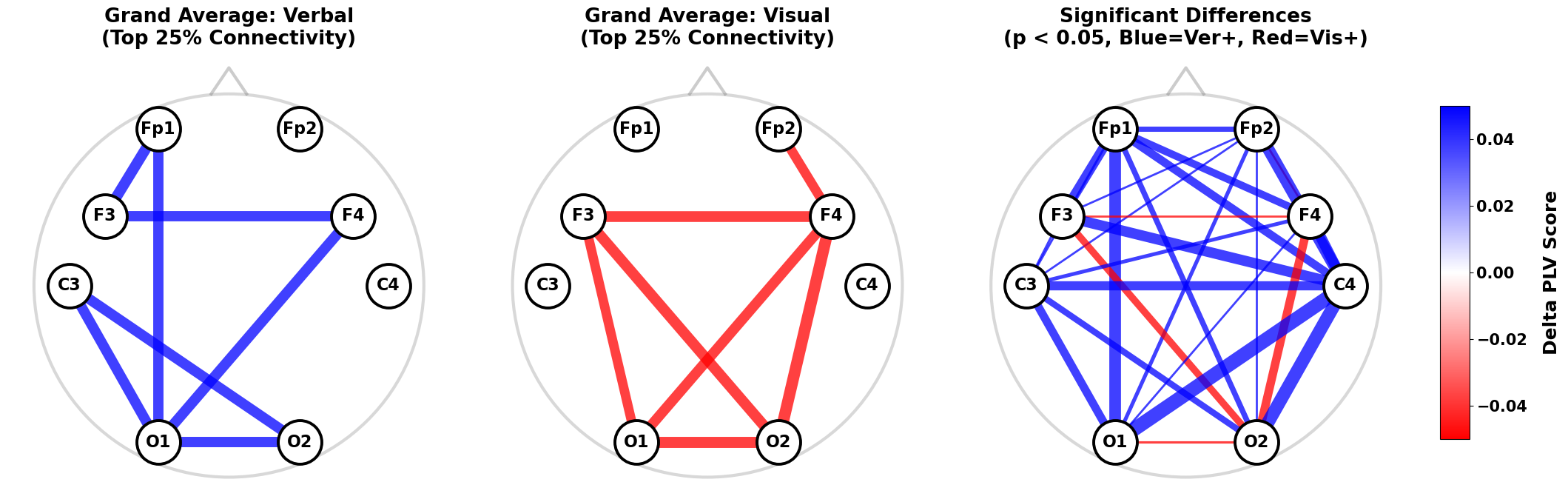}
\caption{Connectivity analysis for the VV dimension: (a) grand average Verbal, (b) grand average Visual, and (c) significant differences.}
\label{fig5}
\end{figure*}

Neural signatures (Fig.~\ref{fig5}) demonstrate clear spatial polarization. The Verbal style forms a dense hub centered in Left-Frontal and Prefrontal regions (Fp1, F3) associated with linguistic circuits and verbal rehearsal \cite{ref17,ref18}. Conversely, the Visual style activates localized Right-Hemisphere and posterior occipital pathways (Fp2-O2, F4-O2) tailored for spatial information processing \cite{ref18}. Significant differences (Fig.~\ref{fig5}c) confirm an overwhelming dominance of red edges (Verbal $>$ Visual). This pronounced, distinct spatial divergence (Anterior vs. Posterior) directly accounts for the superior subject-level classification accuracy (70.00\%) in the VV dimension compared to the overlapping functional networks of AR.

\section{Discussion}

The Leave-One-Subject-Out (LOSO) evaluation reveals a prominent ``Systematic Bias'' rather than a stochastic tracking failure across subject-independent distributions. As demonstrated in the individual voting analysis (Table~\ref{tab4}), several subjects achieved a 0\% classification accuracy despite maintaining highly decisive voting margins (e.g., 20--0). Within the Active--Reflective (AR) dimension, multiple Reflective participants (such as Subjects 11, 12, 13, and 18) were classified as Active with 100\% certainty. This empirical outcome uncovers a distinct ``Systematic Neural Inversion'' effect, indicating that while functional connectivity signatures remain highly stable within individuals, their biological manifestation can be mathematically oriented in a direction diametrically opposed to the population-averaged global model boundary. Similarly, the Verbal--Visual (VV) dimension exhibits a clear binary ``Correct or Inverted'' split (e.g., Subject 8), confirming that while Phase Locking Value (PLV) acts as a highly structured biomarker, its interpretation remains strictly person-specific.

These pronounced 20--0 voting errors provide critical empirical insights into the explicit limitations of utilizing a rigid, ``one-size-fits-all'' global classifier on volatile biological data. A generalized framework inherently presumes a homogeneous distribution of feature representations across an entire population; however, human neurophysiology sharply violates this assumption due to high inter-subject heterogeneity \cite{ref19}. Because the absolute consistency of the misclassifications is not driven by random sensor noise, it confirms that while a population-averaged model can successfully map prominent spatially polarized networks---such as the anterior-posterior divergence during Verbal--Visual processing---its generalizability remains strictly bounded by individual biological diversity. This limitation is heavily reflected in the substantial ``generalization gap'' where perfect intra-subject learnability (100.00\% in the 70:30 split) drops significantly under cross-subject validation, particularly in the AR dimension, where the cross-subject PLV accuracy is limited to 55.56\% due to topological overlaps in shared executive circuits \cite{ref17,ref18}.

\begin{table}[!t]
\caption{Individual Voting Analysis (Selected Subjects)}
\label{tab4}
\centering
\footnotesize
\begin{tabular}{@{}llllcc@{}}
\toprule
\textbf{Dim.} & \textbf{Subj.} & \textbf{True Label} & \textbf{Acc.} & \textbf{V1} & \textbf{V2} \\
\midrule
AR & 3  & Active     & 100\% & 20 & 0 \\
AR & 11 & Reflective & 0\%   & 20 & 0 \\
VV & 1  & Verbal     & 100\% & 20 & 0 \\
VV & 8  & Visual     & 0\%   & 20 & 0 \\
\bottomrule
\end{tabular}
\end{table}

From a methodological perspective, choosing a binary classification paradigm in this preliminary study serves as a critical first step to mathematically isolate these raw discriminative boundaries without over-parameterizing the model. Attempting to map high-dimensional phase-synchronization features to a continuous psychometric scale or a multi-class structure requires a substantially larger cohort to prevent severe overfitting. Given the current sample size ($N=28$), the binary paradigm effectively restricts the parameter space of the SVM classifier, ensuring that the documented neural inversions are genuine representations of neurophysiological variance rather than statistical artifacts of overfitting. Furthermore, these findings underscore that standard subject-wise Z-score normalization remains insufficient to fully eliminate individual neural fingerprints \cite{ref20}. Consequently, subsequent stages of this research framework will focus on structural methodological advancements specifically geared toward mitigating the cross-subject generalization gap. Future work will explore more robust distribution alignment strategies and adaptive feature transformation techniques to better harmonize the heterogeneous neural patterns across different individuals before expanding the scope to systematically incorporate the remaining Sensing--Intuitive and Sequential--Global dimensions of the FSLSM.

\section{Conclusion}

This paper demonstrates that PLV serves as an effective quantitative biomarker for FSLSM recognition. Under LOSO-CV, the VV dimension achieved 70.00\% accuracy due to distinct fronto-occipital polarization, whereas the AR dimension dropped to 55.56\% due to topologically overlapping executive networks. A major contribution of this foundational study is the empirical evaluation of a rigid ``one-size-fits-all'' model, highlighted by the identification of the ``Systematic Neural Inversion'' phenomenon. The presence of high-confidence yet completely inverted predictions (20--0 voting margins) reveals that stable, individual functional connectivity signatures can operate in direct opposition to population-averaged boundaries. Ultimately, these outcomes establish a critical neurophysiological baseline, demonstrating that global models remain strictly bounded by individual biological diversity. These findings strongly motivate future methodological shifts toward adaptive feature transformation techniques specifically designed to mitigate the cross-subject generalization gap.

\section*{Acknowledgment}

This study was funded by the Indonesia Endowment Fund for Education (LPDP) under grant number SKPB-4803/LPDP/LPDP.3/2025.

\IEEEtriggeratref{10}
\bibliographystyle{IEEEtran}
\bibliography{references}

\end{document}